\documentclass{DISproc}

\begin{document}
\title{Charmonium Production in pp Collisions with ALICE}

\author{{\slshape Ionut-Cristian Arsene$^1$ for the ALICE Collaboration}\\[1ex]
{\it $^1$ Research Division and ExtreMe Matter Institute EMMI, GSI Helmholtzzentrum f\"{u}r 
    Schwerionenforschung, Darmstadt, Germany}}



\maketitle


\begin{abstract}
  The ALICE Collaboration studies the production of $J/\psi$ meson in pp
collisions at the center-of-mass energies of 2.76 and \unit{7}{\TeV} at
mid- and forward-rapidity.
  Inclusive production cross-sections are presented as a function
of the collision energy, rapidity and transverse-momentum. 
  The $J/\psi$ polarization measurements in the helicity and
Collins-Soper frames is discussed. 
  A novel result on the correlation between the collision charged particle
multiplicity and $J/\psi$ yield is also shown.
\end{abstract}

\section{Introduction}
Due to their large mass, the heavy quark pairs are considered to be produced in 
hard scatterings of partons which can be described perturbatively.
However, the bound states of heavy quark pairs 
are formed via soft non-perturbative processes. 
Because of this interplay between the perturbative and non-perturbative aspects, 
quarkonium production is a unique and a very important testing case for QCD.
Various theoretical approaches, recently reviewed 
in \cite{reviewBrambilla,reviewLansberg} were proposed to describe the data. However 
the consistent description of both the differential production cross-sections and the polarization
proved to be difficult to achieve.

\begin{wraptable}{l}{0.40\textwidth}
  \centering
  \begin{tabular}{c|c|c}
    \toprule
    & \unit{2.76}{\TeV} & \unit{7}{\TeV} \\
    \midrule
    $|y|<0.9$ & \unit{1.1}{\invnb} & \unit{5.6}{\invnb} \\
    $2.5<y<4$ & \unit{19.9}{\invnb} & \unit{15.6}{\invnb} \\
    \bottomrule
  \end{tabular}
  \caption{Integrated luminosity, $L_{int}$, used in the data analysis at
mid- and forward-rapidity.}
  \label{tab:lumi}
\end{wraptable}

\section{Data analysis}
ALICE \cite{aliceExp} studied the production of $J/\psi$ mesons down to zero transverse-momentum, $p_t$, 
using their
decays into $e^{+}e^{-}$ at mid-rapidity ($|y|<0.9$) and into $\mu^{+}\mu^{-}$ at forward-rapidity ($2.5<y<4$).
In this report we present results on the $J/\psi$ production in pp collisions
at $\sqrt{s}= $\unit{2.76}{\TeV} and \unit{7}{\TeV}. The integrated luminosities of the analyzed data samples
at the two different rapidity intervals are given in Table~\ref{tab:lumi}.
A detailed description of the analysis and the detectors used for reconstructing the electron and muon
candidates can be found in \cite{jpsi7TeV,jpsi276TeV}.
The electrons and muons passing the analysis cuts are combined in opposite-sign (OS) pairs to
construct an invariant mass distribution. In the di-electron channel (at mid-rapidity) the signal
is obtained by subtracting the background which is estimated using the like-sign (LS) pairs \cite{jpsi7TeV}
or track rotations \cite{jpsiMult}.
In the di-muon channel (at forward-rapidity) the signal shape is described by a Crystal Ball function
while the background is parameterized using the sum of two exponentials \cite{jpsi7TeV}. 

In order to extract cross-sections, the raw signal counts extracted from the invariant mass distribution need
to be corrected for triggering efficiencies, kinematical acceptance and reconstruction efficiencies.
This is performed using a Monte-Carlo procedure based on generating a large sample of $J/\psi$ mesons
embedded in simulated pp events. All the particles are then transported through the realistic ALICE detector
setup constructed in GEANT \cite{geant}.

\section{Results}  
\begin{figure}[h]
  \centering
  \begin{tabular}[htb]{cc}
    \includegraphics[width=0.40\textwidth]{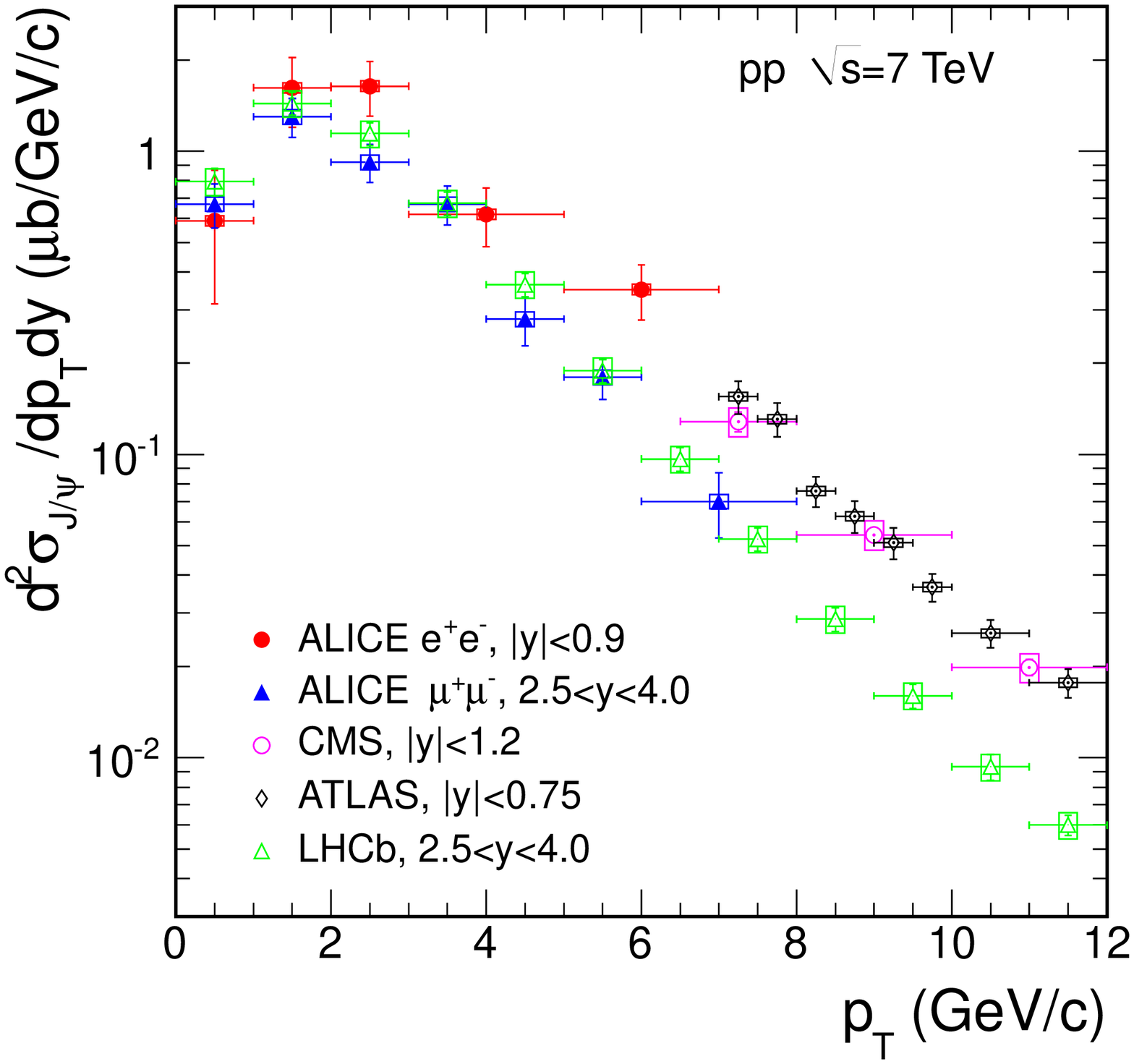}
    &
    \includegraphics[width=0.40\textwidth]{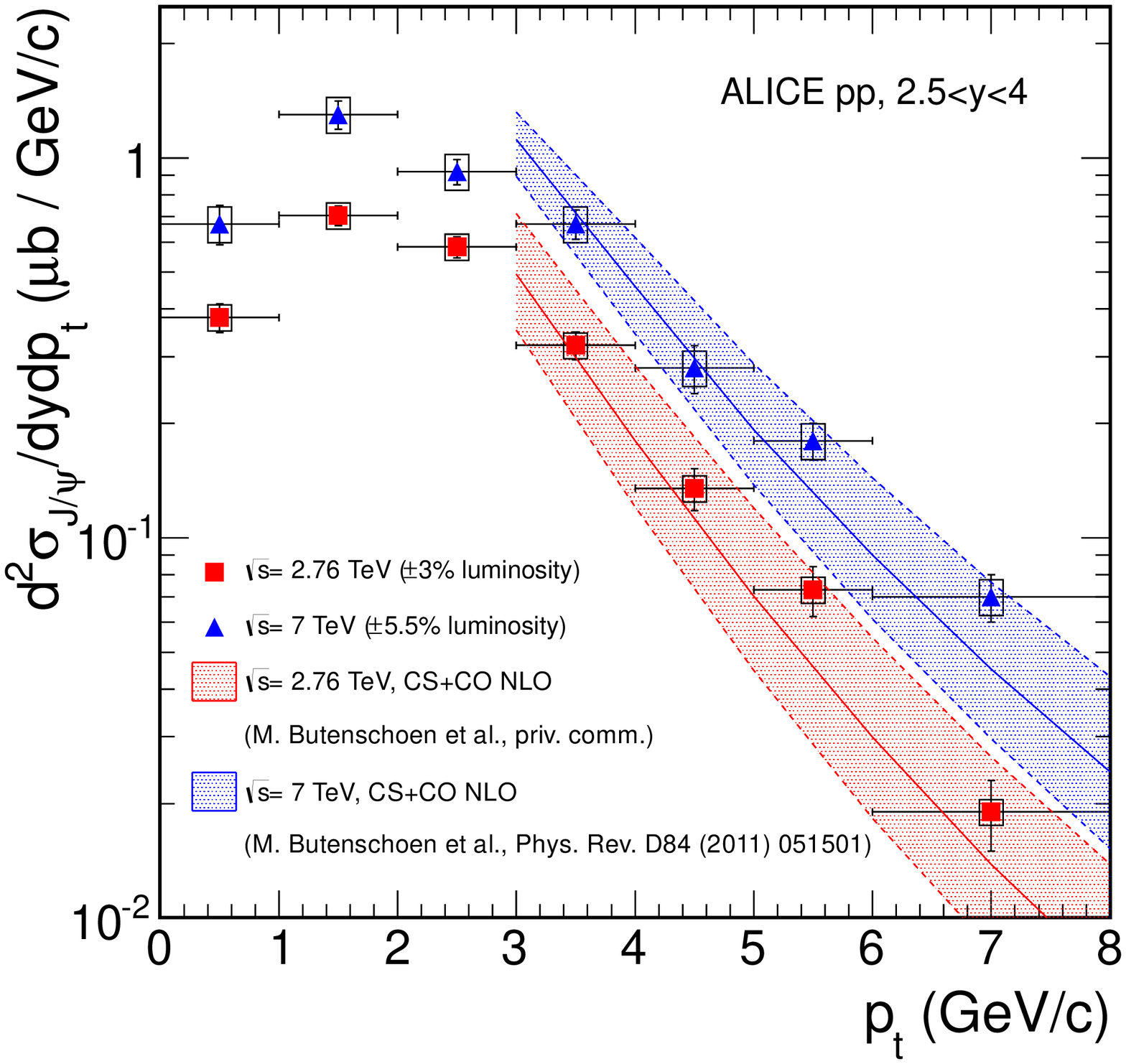} \\
  \end{tabular}
  \caption{Differential cross-section $d^{2}\sigma_{J/\psi}/dp_{t}dy$ in pp collisions at 
           \unit{7}{\TeV} \cite{jpsi7TeV} and \unit{2.76}{\TeV} \cite{jpsi7TeV}.
           The model calculations are from \cite{kniehl}.}
  \label{Fig:dndpt}
\end{figure}

Figure~\ref{Fig:dndpt} presents the inclusive differential cross-section $d^{2}\sigma_{J/\psi}/dp_{t}dy$
in pp collisions at $\sqrt{s}=$\unit{7}{\TeV} \cite{jpsi7TeV} and 
\unit{2.76}{\TeV} \cite{jpsi276TeV}. At \unit{7}{\TeV} the $p_t$ dependent cross-sections
at mid-rapidity and forward-rapidity are shown together with the results from CMS \cite{cmsPt} 
and ATLAS \cite{atlasPt} at mid-rapidity and LHCb \cite{lhcbPt} at forward-rapidity.
At mid-rapidity, the $p_t$ coverage of the ALICE results is complementary to that from CMS and ATLAS.
At forward-rapidity the ALICE results are in agreement with those from LHCb.

The inclusive cross-section at forward-rapidity in pp collisions at $\sqrt{s}=$\unit{2.76}{\TeV}
and \unit{7}{\TeV} is shown in the right panel of Figure~\ref{Fig:dndpt}.
The results at both energies are compared with the predictions of a NRQCD calculation \cite{kniehl}
which includes both color singlet and color octet terms at NLO.

\begin{figure}[htb]
  \centering
  \begin{tabular}[htb]{cc}
    \includegraphics[width=0.40\textwidth]{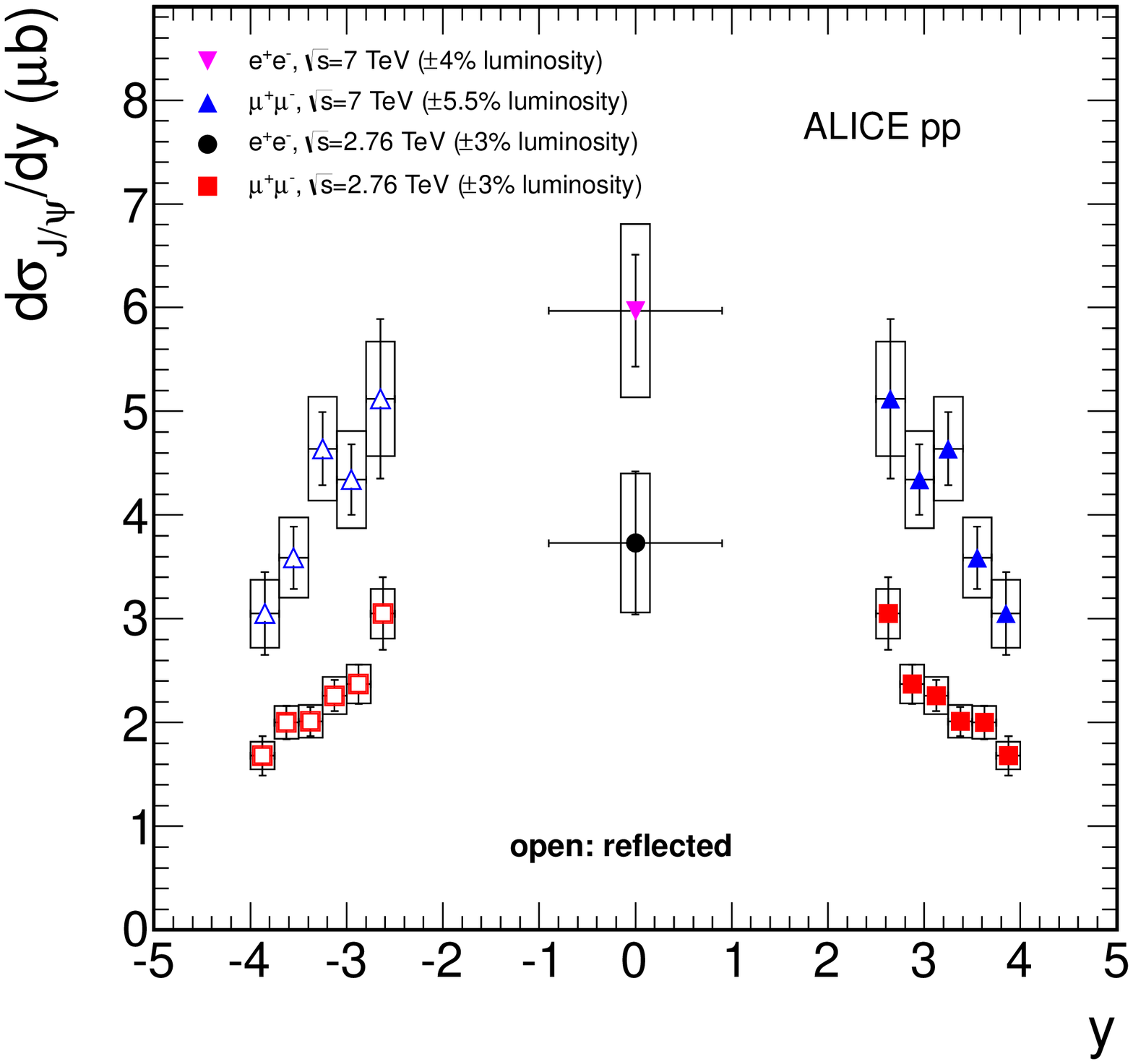}
    &
    \includegraphics[width=0.40\textwidth]{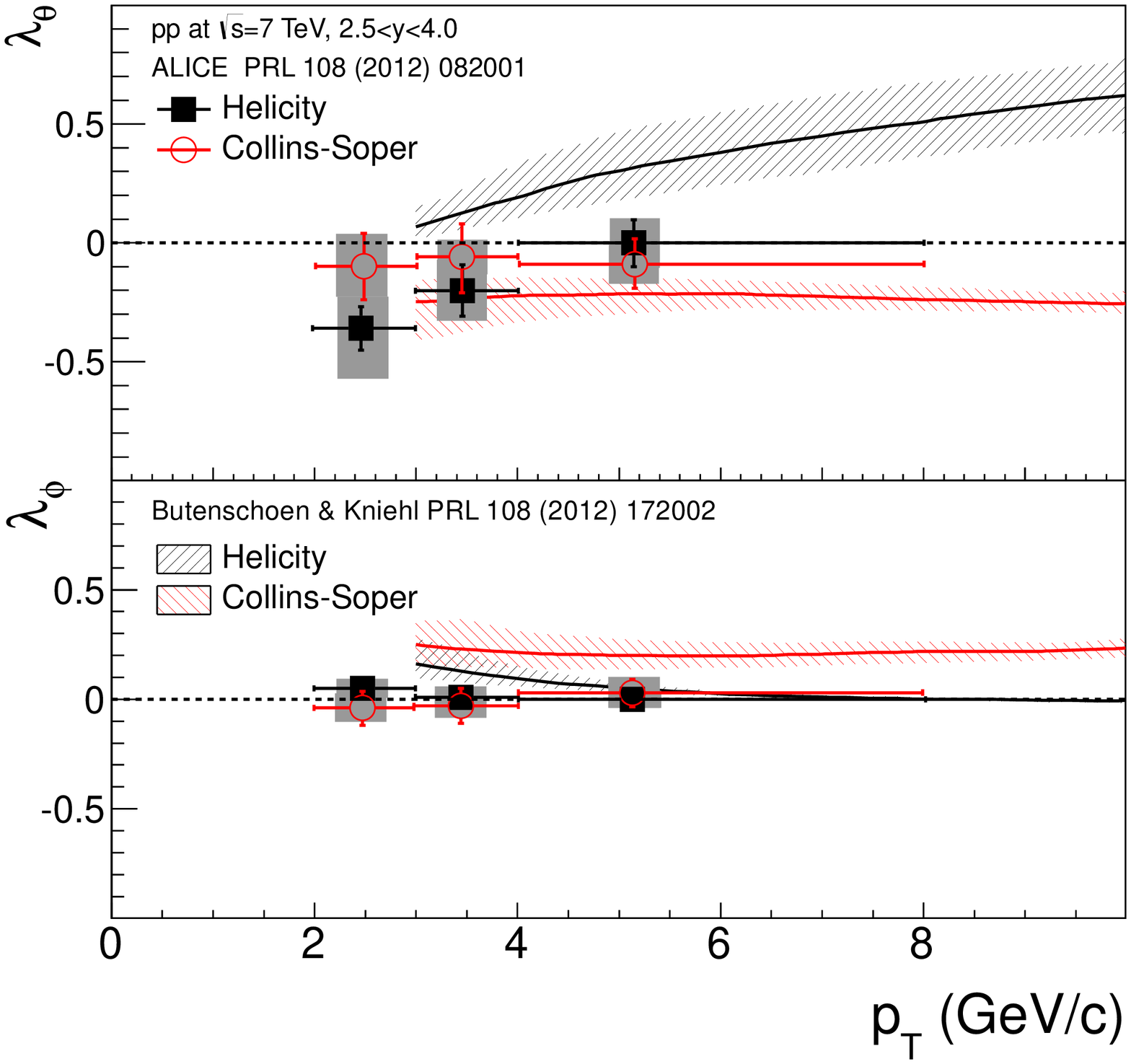} \\
  \end{tabular}
  \caption{Left: $d\sigma_{J/\psi}/dy$ in pp collisions \cite{jpsi276TeV}.
           Right: Polarization parameters $\lambda_{\theta}$ and $\lambda_{\phi}$ as a function
           of $p_t$ for inclusive $J/\psi$ measured in the helicity and Collins-Soper frames \cite{jpsiPolariz}.}
  \label{Fig:dndyPolariz}
\end{figure}

The wide kinematic coverage of ALICE, which is unique among the LHC experiments, allows the extraction of 
the $p_t$ integrated $J/\psi$ cross-section. In the left panel of Figure~\ref{Fig:dndyPolariz} we present 
the $d\sigma_{J/\psi}/dy$ at $\sqrt{s}=$2.76 and \unit{7}{\TeV}. The results refer to the inclusive
$J/\psi$ production which is a sum of the direct component and of $J/\psi$ resulting
from decays of higher-mass charmonium states (mainly the $\chi_{c1}$, $\chi_{c2}$ and $\Psi(2S)$ states)
and from b-hadron decays. The contribution from higher-mass charmonium states measured at 
lower energies \cite{faciolli,phenix} amounts to $\approx$33\% while the contribution from
b-hadron decays is 10-15\% in the $p_t$ range covered by ALICE \cite{lhcbPt,jpsiB}.

The $J/\psi$ is a spin-1 boson allowing for three degenerated states corresponding to 
projections of the angular momentum $J_z=\pm1$ (transversal polarization) and $J_z=0$ (longitudinal
polarization). 
\begin{wrapfigure}{r}{0.40\textwidth}
  \centering
  \includegraphics[width=0.40\textwidth]{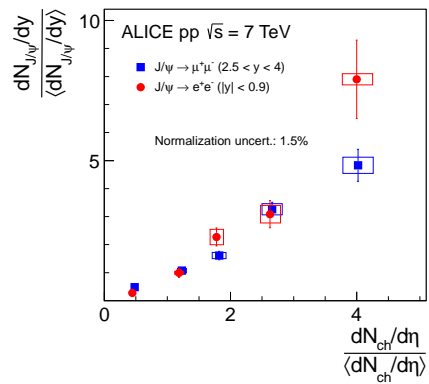}
  \caption{$J/\psi$ yield as a function of the charged particle density at mid-rapidity \cite{jpsiMult}.}
  \label{Fig:jpsiMult}
\end{wrapfigure}
The observed polarization is a superposition of the polarization from all the
production mechanisms thus making this measurement a very important constraint for theoretical
calculations. 
Existing models have difficulties in describing at the same time
both the $J/\psi$ production cross-section and the polarization. 
In particular NRQCD at leading order predicts for high-$p_t$ $J/\psi$ a large
transverse polarization at CDF energies \cite{CDFpolariz}.

ALICE measured the $J/\psi$ polarization at forward-rapidity in the helicity ($z-$axis is the $J/\psi$ 
own momentum direction
in the center-of-mass frame of the pp collision) and Collins-Soper ($z-$axis is the bisector of the angle
between the direction of one beam and the opposite of the direction of the other one, in the rest frame
of the $J/\psi$) frames.
The measured angular distribution of the decay leptons is parameterized using the 
general form \cite{faciolliPolariz}
\begin{align}
W(\theta,\phi) \approx \frac{1}{3+\lambda_{\theta}}(1+\lambda_{\theta}\mathrm{cos}^2\theta + 
                 \lambda_{\phi}\mathrm{sin}^2\theta \mathrm{cos}2\phi
               +\lambda_{\theta\phi}\mathrm{sin}2\theta \mathrm{cos}\phi),\nonumber
  \label{eq:polariz}
\end{align}
where $\theta$ ($\phi$) are the polar (azimuthal) angles. The $\lambda_{\theta}$, $\lambda_{\phi}$ and
$\lambda_{\theta\phi}$ are parameters extracted from data which quantify the degree of polarization. 
The right panel of Figure~\ref{Fig:dndyPolariz} shows the ALICE results on $\lambda_{\theta}$ and $\lambda_{\phi}$ 
for inclusive $J/\psi$ at forward-rapidity \cite{jpsiPolariz}.
In both reference frames all the parameters are compatible with zero. Recent NLO calculations
within the NRQCD factorization \cite{kniehlPolariz} have shown good agreement with the ALICE results.

To investigate further the $J/\psi$ production mechanisms, the yield was measured as a function of
the charged particle pseudo-rapidity density $dN_{ch}/d\eta$.
Figure~\ref{Fig:jpsiMult} presents the relative $J/\psi$ yield at mid- and forward-rapidity as a function 
of the relative charged particle density at mid-rapidity \cite{jpsiMult}. 
The results indicate that the $J/\psi$ production at both mid- and 
forward-rapidity tends to be accompanied by the production of many other charged hadrons. 
A possible reason for the observed results could be multiple partonic interactions \cite{strikman,fereiro}.

\section{Conclusions}
We presented results obtained by the ALICE Collaboration on $J/\psi$ production in pp collisions 
at $\sqrt{s}=$2.76 and \unit{7}{\TeV}. The inclusive cross-sections as a function of $p_t$ and rapidity
were shown. The NLO NRQCD calculations show a good agreement with the ALICE results at forward-rapidity.
The measured polarization parameters $\lambda_{\theta}$ and $\lambda_{\phi}$ are compatible
with zero. We have also shown that the relative $J/\psi$ yields at mid- and forward-rapidity increase linearly
with the charged particle density at mid-rapidity.

\section{Acknowledgement}
This work was supported by the Helmholtz Alliance Program of the Helmholtz Association, contract
HA216/EMMI "Extremes of Density and Temperature: Cosmic Matter in the Laboratory".

{\raggedright
\begin{footnotesize}

\end{footnotesize}
}

\end{document}